\documentstyle[12pt]{article}
\def\inbar{\,\vrule height1.5ex width.4pt depth0pt}
\def\I1{\relax\hbox{$\inbar\kern-.35em{\rm 1}$}}
 \font\cmss=cmss10 \font\cmsss=cmss10 at 7pt

\def\IZ{\relax\ifmmode\mathchoice
 {\hbox{\cmss Z\kern-.4em Z}}{\hbox{\cmss Z\kern-.4em Z}}
 {\lower.9pt\hbox{\cmsss Z\kern-.4em Z}}
 {\lower1.2pt\hbox{\cmsss Z\kern-.4em Z}}
 \else{\cmss Z\kern-.4em Z}\fi}

\begin{document}

\hyphenation{Schwarz-schild}
\hyphenation{Lind-quist}
\hyphenation{di-men-sion-al}

\thispagestyle{empty}
\rightline{PUPT-1548}
\rightline{hep-th/9506200}
\rightline{June 1995}

\vskip2cm
\begin{center}
{\large\bf Entropy and supersymmetry of $D$ dimensional \break
extremal electric black holes versus string states} \break

\vskip 1 cm
{\bf Amanda W. Peet}
\footnote{Electronic address: peet@puhep1.princeton.edu},

Joseph Henry Laboratories, \break
Princeton University,  \break
Princeton, NJ 08544, U.S.A.

\end{center}
\vskip1cm

\begin{quote}

Following the work of Sen, we consider the correspondence between
extremal black holes and string states in the context of the entropy.
We obtain and study properties of electrically charged black hole
backgrounds of tree level heterotic string theory compactified on a
$p$ dimensional torus, for $D=(10-p)=4 \ldots 9$.  We study in
particular a one--parameter extremal class of these black holes, the
members of which are shown to be supersymmetric.  We find that the
entropy of such an extremal black hole, when calculated at the stringy
stretched horizon, scales in such a way that it can be identified with
the entropy of the elementary string state with the corresponding
quantum numbers.

\end{quote}
\vskip1cm

\normalsize
\newpage


\section*{Introduction}

Recently there has been considerable interest in the question of
whether black hole configurations of string theory can or must be
identified as string states.  The question has arisen in studies
of duality, of the configurations themselves, and also in the
study of the black hole information problem
\cite{S1,Vafa,RS,DuffRahmfeld,DuffetalReview,HullTownsend,Witten,Sen1}
{}.

Qualitatively, the idea of identifying string states as black
holes is a simple one.  At a given value of the string coupling,
most states above a certain mass level have a Compton wavelength
which is shorter than their Schwarzschild radius, and so may be
considered to be black holes.  However, there are some string
states which should not be identified as black holes, because
they rotate too fast for their effective radius to be inside
their Schwarzschild radius.

The question of whether nonextremal uncharged black holes could
be identified as string states was considered in the context of
the entropy in \cite{S1,RS}.  The entropy for the black holes was
obtained using the usual formula involving the area of the event
horizon in the Einstein metric, and for the string states it was
simply the logarithm of the degeneracy of the string states with
a given mass (and angular momentum).  It was found that the
scaling permitted an identification of black holes as string
states, as long as the mass used for the black hole was the
Rindler mass at the stretched horizon\cite{S1}.  This Rindler
mass is related to the {\sc ADM} mass by the redshift between the
stretched horizon and asymptotic infinity, which is a large
number of order the mass of the black hole.  The stretched
horizon, located at a proper distance of order one string unit
away from the event horizon, is the place where stringy effects
start coming into play which are likely to be important for the
resolution of the black hole information problem
\cite{Scollab,STU,SU}.

For extremal black holes, a comparison with string states was
made in the context of heterotic string theory compactified on a
six dimensional torus\cite{DuffRahmfeld}.  It was found that the
quantum numbers of several extremal black hole solutions matched
those of supersymmetric string states of the same theory.  Then
in a very interesting work\cite{Sen1}, it was found that the
entropy of extremal electrically charged black holes, when
calculated at the stringy stretched horizon, scaled in such a way
that it agreed with the string entropy obtained by taking the
logarithm of the degeneracy of string states with corresponding
quantum numbers.  It was necessary in this case to calculate the
entropy at the stringy stretched horizon: classically the area of
the event horizon of these extremal black holes is zero, but this
is not expected to survive quantum corrections.  For extremal
black holes there is no redshift--induced renormalization of the
mass, and so the agreement of the entropies was striking.

In \cite{Vafa} a concern had been voiced that the scaling of the
entropy for black holes in $D$ dimensions might preclude an
identification with string states.  However, it was seen in the
abovementioned works that this turned out not to be a problem in
$D=4$, provided that one calculated the relevant quantities at
the stretched horizon.

In this work we will study electrically charged black hole
backgrounds of tree level heterotic string theory compactified on
a $p$ dimensional torus, where $D=(10-p) = 4\ldots 9$, with a
view to understanding whether the correspondence found in
\cite{Sen1} is specific to $D=4$.  Along the way, we study
supersymmetry, which is expected\cite{Sen3} to exist for the
extremal black holes in $D=4$ which have a fixed relation between
{\sc ADM} mass and right--handed charge.

The plan of the paper is as follows.  In Section 1 we review the
solution--generating techniques which will be used in Section 2
to obtain the $D$ dimensional electrically charged black holes.
Supersymmetry of a one--parameter extremal ``type $\cal{R}$''
class of these black holes is studied in Section 3, where it is
found that half of the possible supersymmetries are unbroken.
Section 4 contains a discussion of the entropy and the stretched
horizon of these black holes.  We find that the entropy of type
$\cal{R}$ black holes, when calculated at the stringy stretched
horizon, agrees with the entropy of elementary string states of
the same quantum numbers.  We end with some conclusions; our
notation and conventions are listed in the appendix.


\section{Actions and solution--generating}
\setcounter{equation}{0}

In order to obtain general spherically symmetric electrically
charged black hole backgrounds of the tree level action for
heterotic string theory compactified on a $p$ dimensional torus,
we will employ solution--generating techniques.  We will begin by
reviewing the tree level actions in $D$ and ten dimensions, and
the solution--generating methods\cite{SolGen}.

The massless bosonic fields in heterotic string theory
compactified on a $p$ dimensional torus are the metric tensor
$G_{\mu\nu}$, the anti-symmetric tensor field $B_{\mu\nu}$,
$k=(16+2p)$ $U(1)$ gauge fields $A_\mu^{\alpha}$ ($1\le\alpha\le
k$), the scalar dilaton field $\Phi$, and a $k\times k$
matrix--valued scalar moduli field $M$ satisfying
\begin{equation}
M L M^T = L \qquad M^T = M.
\end{equation}
In this equation $L$ is a $k\times k$ symmetric matrix with
$(16+p)$ eigenvalues $-1$ and $p$ eigenvalues $+1$.  We will take
$L$ to be
\begin{equation}
L=\pmatrix{-\I1_{16+p} & \cr & \I1_p\cr}
\end{equation}
where $\I1_n$ denotes the $n\times n$ identity matrix.

The action describing the effective field theory of these fields
is \cite{MaharanaSchwarz,Sen2}
\begin{eqnarray}\label{stringaction}
S &=& \int d^D x \sqrt{-G} \, e^{-\Phi} \, \Big[ R[G]
+ G^{\mu\nu} \partial_\mu \Phi \partial_\nu\Phi
+\textstyle\frac{1}{8} G^{\mu\nu}
{\mbox{Tr}} (\partial_\mu M L\partial_\nu ML)
\nonumber \\
&& -\textstyle\frac{1}{12}
G^{\mu\lambda} G^{\nu\kappa} G^{\rho\sigma} H_{\mu\nu\rho}
H_{\lambda\kappa\sigma} - \textstyle\frac{1}{4}
G^{\mu\lambda} G^{\nu\kappa} F^{\alpha}_{\mu\nu} \,
(LML)_{\alpha\beta} \, F^{\beta}_{\lambda\kappa} \Big]
\end{eqnarray}
where
\begin{eqnarray}
F^{\alpha}_{\mu\nu} &=& \partial_\mu A^{\alpha}_\nu -
\partial_\nu A^{\alpha}_\mu
\nonumber\\
H_{\mu\nu\rho} &=& \partial_\mu B_{\nu\rho}
+ \textstyle\frac{1}{2}
A_\mu^{\alpha} L_{\alpha\beta} F^{\beta}_{\nu\rho}
+(\hbox{cyclic})
\end{eqnarray}
and $R[G]$ is the scalar curvature formed from the metric
$G_{\mu\nu}$.

We note that it is $e^{-\Phi}$ which appears in front of the
action in string metric (\ref{stringaction}), so we have for the
string coupling at asymptotic infinity
\begin{equation}
g_\infty = \langle e^{\Phi/2} \rangle_\infty \quad .
\end{equation}

The action (\ref{stringaction}) is invariant under the
$O(16+p,p)$ transformation
\begin{equation}
M\to \Omega M \Omega^T \qquad A_\mu \to \Omega A_\mu \qquad
\Phi\to \Phi \qquad G_{\mu\nu}\to G_{\mu\nu} \qquad B_{\mu\nu}
\to B_{\mu\nu}
\end{equation}
where $\Omega$ is a $k\times k$ matrix satisfying
\begin{equation}
\Omega L \Omega^T = L\quad .
\end{equation}
This is a symmetry of the full string theory if the $k$
dimensional lattice $\Lambda$ of electric charges is also rotated
by $\Omega$ \cite{Narain}.  For fixed $\Lambda$, only a discrete
subgroup $O(16+p,p;\IZ )$ is a symmetry.

For time--independent backgrounds, it is expected that there is
an enlarged symmetry, namely $O(17+p,1+p)$ \cite{HassanSen}.  In
the barred variables, defined by
\begin{eqnarray}\label{abgphibars}
\bar A^{\alpha}_i &=& A^{\alpha}_i - (G_{tt})^{-1} G_{ti}
A^{\alpha}_t \quad\quad 1\le\alpha\le k \quad 1\le i\le (D-1)
\nonumber \\
\bar A^{(k+1)}_i &=& (G_{tt})^{-1} G_{ti}
\nonumber \\
\bar A^{(k+2)}_i &=& B_{it} + \textstyle\frac{1}{2}
A^{\alpha}_t L_{\alpha\beta} \bar A^{\beta}_i
\nonumber\\
\bar G_{ij} &=& G_{ij} - (G_{tt})^{-1} G_{ti} G_{tj}
\nonumber\\
\bar B_{ij} &=& B_{ij} + \bar A_{[i}^{(k+1)}\bar A_{j]}^{(k+2)}
\nonumber\\
\bar \Phi &=& \Phi - \textstyle\frac{1}{2} \ln (-G_{tt})
\end{eqnarray}
and
\begin{eqnarray}\label{mbar}
\bar M^{-1} &=& \pmatrix{M + (G_{tt})^{-1} A_t A_t^T & -
(G_{tt})^{-1} A_t & ML A_t + (G_{tt})^{-1} A_t (A_t^T L A_t)\cr &
& \cr - (G_{tt})^{-1} A^T_t & (G_{tt})^{-1} & -(G_{tt})^{-1}
A_t^T L A_t\cr &{}&\cr A_t^TLM & -(G_{tt})^{-1} A_t^T L A_t &
G_{tt} + A_t^T LML A_t \cr + (G_{tt})^{-1} A_t^T (A_t^TLA_t) &{}&
+ (G_{tt})^{-1} (A^T_t L A_t)^2\cr} \nonumber\\
& &
\end{eqnarray}
and
\begin{equation}
\bar L = \pmatrix{ L &0 &0 \cr 0&0& 1 \cr 0&1 &0 \cr}
\end{equation}
the action can be rewritten as
\begin{eqnarray}
S &=& C\int dt \int d^{D-1}x \sqrt{\bar G} e^{-\bar \Phi} \Big[
R_{\bar G} + \bar G^{ij} \partial_i \bar\Phi \partial_j\bar \Phi
+\textstyle\frac{1}{8} \bar G^{ij} Tr(\partial_i \bar M \, \bar
L\partial_j \bar M \, \bar L) \nonumber \\ &&
-\textstyle\frac{1}{12}\bar G^{il}\bar G^{jm} \bar G^{kn} \bar
H_{ijk} \bar H_{lmn} - \textstyle\frac{1}{4} \bar G^{ik} \bar
G^{jl} \bar F^{\bar\alpha}_{ij} (\bar L\bar M \bar
L)_{\bar\alpha\bar\beta}\bar F^{\bar\beta}_{kl} \Big]
\end{eqnarray}
where
\begin{eqnarray}
\bar F^{\bar\alpha}_{ij} &=& \partial_i \bar A^{\bar\alpha}_j -
\partial_j \bar A^{\bar\alpha}_i \, , \qquad 1\le \bar\alpha \le
k+2, \nonumber\\ \bar H_{ijk} &=& \partial_i \bar B_{jk} +
\textstyle\frac{1}{2} \bar A_i^{\bar\alpha} \bar
L_{\bar\alpha\bar\beta} \bar F^{\bar\beta}_{jk}
+\hbox{(cyclic)}\quad .
\end{eqnarray}
The $O(17+p,1+p)$ symmetry then acts as
\begin{equation}\label{o1p17p}
\bar M\to \bar \Omega \bar M \bar \Omega^T \qquad \bar A_i \to
\bar \Omega \bar A_i \qquad \bar\Phi\to \bar \Phi \qquad \bar
G_{ij}\to\bar G_{ij} \qquad \bar B_{ij} \to \bar B_{ij}
\end{equation}
where $\bar\Omega$ is a $(k+2)\times (k+2)$ matrix satisfying
\begin{equation}
\bar \Omega \bar L \bar \Omega^T = \bar L\quad .
\end{equation}

We will work with the parametrization of $\bar\Omega$ where $\bar
L$ is diagonal; the orthogonal matrix $U$ that diagonalizes $\bar
L$ is given by
\begin{equation}
U=\pmatrix{ \I1_{k} && \cr & \frac{+1}{\sqrt{2}} &
\frac{+1}{\sqrt{2}} \cr & \frac{-1}{\sqrt{2}} &
\frac{+1}{\sqrt{2}} \cr}
\end{equation}
so that
\begin{equation}
U \bar L U^T \equiv \bar L_d = {\mbox{diag}}( -\I1_{16+p} \, , \,
\I1_p \, , \, 1 \, , \, -1) \quad .
\end{equation}
Then $U \bar \Omega U^T$ preserves $\bar L_d$.

As was done in \cite{Sen3}, we will apply this $O(17+p,1+p)$
transformation to a known time--independent classical black hole
solution to generate more classical solutions of the equations of
motion.  Without loss of generality, we can restrict to solutions
with fields which tend to zero at asymptotic infinity -- except
for the metric, which tends to the Minkowski metric, and the
moduli matrix which tends to the unit matrix.  With these
conditions, we use only the subgroup $O(16+p,1)~\otimes~O(p,1)$;
the rest of the group generates pure gauge
deformations\cite{HassanSen}.  The solution which we start with
will be uncharged and will have a trivial (unit) moduli matrix,
so we see by looking at (\ref{abgphibars})--(\ref{mbar}) that we
will need to mod out by an $O(16+p)\otimes O(p)$ symmetry.  Thus
the transformations which generate inequivalent solutions live in
the coset space
\begin{equation}
\bigl[ O(16+p,1)\otimes O(p,1)\bigr]/
       \bigl[(O(16+p)\otimes O(p) \bigr]
\end{equation}
which has dimension $(18+2p)=(k+2)$.  We will use the following
parametrization \cite{Sen3} of a generic element:
\begin{equation}
\bar \Omega = \bar\Omega_2 \bar \Omega_1
\end{equation}
where
\begin{equation}\label{Uomega1U}
U\bar \Omega_1 U^T = \pmatrix{ \I1_{15+p} & 0 & 0 & 0 & 0 & 0 \cr
0 & \cosh\alpha & 0 & 0 & \sinh\alpha & 0 \cr 0 & 0 & \I1_{p-1} &
0 & 0 & 0 \cr 0 & 0 & 0 & \cosh\beta & 0 & \sinh\beta \cr 0 &
\sinh\alpha & 0 & 0 & \cosh\alpha & 0 \cr 0 & 0 & 0 & \sinh\beta
& 0 & \cosh\beta \cr}
\end{equation}
and
\begin{equation}\label{omega2}
\bar \Omega_2 = \pmatrix{R_{16+p}(\vec n_L) && \cr & R_p(\vec
n_R) & \cr && \I1_2\cr}
\end{equation}
where
\begin{equation}
R_N(\vec n) \pmatrix{{\vec{0}_{N-1}}\cr 1} = \vec n \quad .
\end{equation}
Here $\vec n_L$ and $\vec n_R$ are arbitrary $(16+p)$ and $p$
dimensional unit vectors respectively.  So $\bar \Omega$ in the
above equations is parametrized by $(k+2)$ parameters: $\alpha$,
$\beta$, $\vec n_L$ and $\vec n_R$.  $\alpha,\beta \in
[0,\infty)$ are the boost parameters of the noncompact group we
use to generate solutions; $\alpha$ is the parameter of the
left--handed part of the group and $\beta$ is the parameter of
the right--handed part of the group.

We now turn to the question of how ten and $D$ dimensional fields
are related.  The relationship between the vector and moduli
fields and the ten dimensional fields is a little involved
\cite{MaharanaSchwarz}.  The first step of the dictionary is a
change of basis.  In \cite{MaharanaSchwarz,SenIJMP} the matrix
$L$ is chosen to be off--diagonal, and in order to take the
moduli matrix and vector fields to the ones appropriate to the
off--diagonal $L$, the matrix
\begin{equation}\label{basisU}
U_\star = \pmatrix{\I1_{16} & 0 & 0 \cr
0 & {\textstyle{\frac{1}{\sqrt{2}}}}\I1_p &
{\textstyle{\frac{-1}{\sqrt{2}}}}\I1_p \cr
0 & {\textstyle{\frac{1}{\sqrt{2}}}}\I1_p &
{\textstyle{\frac{1}{\sqrt{2}}}}\I1_p \cr}
\end{equation}
must be used to transform $A$ and $M$ as
\begin{equation}\label{basisAM}
A \to A_\star = U_\star^T A \qquad \qquad
M \to M_\star = U_\star^T M U_\star
\end{equation}
where $A=\pmatrix{A_L \cr A_R}$.

Once this change of basis has been done, and with a split of the
left--handed directions into $16$ and $p$ dimensional parts,
\begin{eqnarray}
M_\star &=& \pmatrix{ \I1_{16} + a G^{-1} a^T & a G^{-1} C+a & a
G^{-1} \cr C^T G^{-1} a^T + a^T & G + C^T G^{-1} C + a^T a & C^T
G^{-1} \cr G^{-1} a^T & G^{-1}C & G^{-1} \cr} \qquad A_\star =
\pmatrix{ A^{[3]} \cr A^{[2]} \cr A^{[1]} \cr} \nonumber\\
& &
\end{eqnarray}
where $C=B+{\textstyle{\frac{1}{2}}}a^Ta$, $B=(B_{\alpha\beta})$,
$a=(a_\alpha^I)$, etc.  Then (upon doing a sign change on the
left--handed $A_\star$ fields which is necessary\cite{Sen404} to
convert them to the signature convention of \cite{MaharanaSchwarz}) we
have for the ten dimensional fields\cite{MaharanaSchwarz}, denoted by
tildes,
\begin{eqnarray}\label{tenvsD}
\tilde A^I_\alpha &=& a^I_\alpha \nonumber\\ \tilde A_\mu^I &=&
a^I_\alpha A^{[1]\alpha}_\mu + A^{[3]I}_\mu \nonumber\\ \tilde
B_{\alpha\beta} &=& B_{\alpha\beta} \nonumber\\ \tilde B_{t\alpha} &=&
A^{[2]}_{t\alpha}-B_{\alpha\beta} A_t^{[1]\beta}
-{\textstyle{\frac{1}{2}}}a^I_\alpha A_t^{[3]I} \nonumber\\ \tilde
G_{\alpha\beta} &=& G_{\alpha\beta} \nonumber\\ \tilde E &=&
\pmatrix{E^m_\mu & A_\mu^{[1]\alpha}E^a_\alpha \cr 0 & E^a_\alpha \cr}
\nonumber\\ \tilde E^{-1} &=& \pmatrix{E^\mu_m & -A_\mu^{[1]\alpha}
E^\mu_m \cr 0 & E^\alpha_a \cr} \nonumber\\ \tilde\omega[\tilde
G]_{0,0i} &=& -\partial_i\Phi \qquad\qquad\qquad\qquad\quad
\tilde\omega[\tilde G]_{0,ia} = \textstyle\frac{1}{2}
F^{[1]}_{ti\alpha} E^\alpha_a E^t_0 \nonumber\\ \tilde\omega[\tilde
G]_{i,ab} &=& \textstyle\frac{1}{2} E^\alpha_a \partial_i E_{\alpha b}
- (a \leftrightarrow b) \qquad\quad \tilde\omega[\tilde G]_{i,0a} =
\textstyle\frac{-1}{2} F^{[1]}_{ti\alpha}E^\alpha_a E^t_0 \nonumber\\
\tilde\omega[\tilde G]_{a,0i} &=& \textstyle\frac{-1}{2}
F^{[1]\alpha}_{ti}E^t_0 E_{\alpha a} \qquad\qquad\qquad
\tilde\omega[\tilde G]_{a,ib} = \textstyle\frac{-1}{2} E^\alpha_a
E^\beta_b \partial_i G_{\alpha\beta} \nonumber\\ \tilde\Phi &=& \Phi +
{\log{\sqrt{\det(G_{\alpha\beta})}}}
\end{eqnarray}
where $\tilde\omega$ are the
spin--connections\cite{ScherkSchwarz}.  We have not listed
$\tilde B_{\mu\nu}, B_{\mu\nu}$ because they will be zero for the
black holes which we study.  The above ten versus $D$ dimensional
field relations will be useful in Section 3 where we consider
supersymmetry.  The action of the ten dimensional fields is then
\begin{eqnarray}
S_{10} &=& \int d^{10}x \sqrt{-\tilde G}e^{-\tilde\Phi}\biggl[
R[\tilde G] + \tilde G^{\tilde\mu\tilde\nu}
\partial_{\tilde\mu}\tilde\Phi \partial_{\tilde\nu}\tilde\Phi
-{\textstyle{\frac{1}{12}}} \tilde G^{\tilde\mu\tilde\tau} \tilde
G^{\tilde\nu\tilde\sigma} \tilde G^{\tilde\rho\tilde\kappa}
\tilde H_{\tilde\mu\tilde\nu\tilde\rho} \tilde
H_{\tilde\tau\tilde\sigma\tilde\kappa} \nonumber\\
& & -{\textstyle{\frac{1}{4}}}
\tilde G^{\tilde\mu\tilde\tau} \tilde G^{\tilde\nu\tilde\sigma}
\tilde F_{\tilde\mu\tilde\nu} \tilde F_{\tilde\tau\tilde\sigma}
\biggr] \quad .
\end{eqnarray}

We end this Section by giving the relationship between string and
Einstein variables.  In $D=(10-p)$ dimensions, the relation
between the string metric $G$ and the Einstein metric $g$ is
\begin{equation}\label{stringeinstein}
G_{\mu\nu} = e^{\gamma\Phi} g_{\mu\nu}
\end{equation}
where we have defined the fraction $\gamma$ to be
\begin{equation}\label{gammadef}
\gamma = \frac{2}{(D-2)} \quad .
\end{equation}
In four dimensions $\gamma=1$.  In the Einstein metric, the $D$
dimensional action may then be expressed as
\begin{eqnarray}\label{einsteinaction}
S[g] &=& \int d^D x \sqrt{- g} \, \Big[ R[g] -
{\textstyle\frac{1}{(D-2)}} g^{\mu\nu} \partial_\mu \Phi
\partial_\nu\Phi +\textstyle\frac{1}{8} g^{\mu\nu} {\mbox{Tr}}
(\partial_\mu M L\partial_\nu ML) \nonumber \\ &&
-\textstyle\frac{1}{12} e^{-2\gamma\Phi} g^{\mu\lambda}
g^{\nu\kappa} g^{\rho\sigma} H_{\mu\nu\rho}
H_{\lambda\kappa\sigma} - {\textstyle{\frac{1}{4}}}
e^{-\gamma\Phi} g^{\mu\lambda} g^{\nu\kappa} F^{\alpha}_{\mu\nu}
\, (LML)_{\alpha\beta} \, F^{\beta}_{\lambda\kappa} \Big] \nonumber\\
& &
\end{eqnarray}


\section{The black holes}
\setcounter{equation}{0}

In this Section we will follow \cite{Sen3} and apply the
solution--generating transformations of the previous Section in
order to obtain electrically charged black hole backgrounds.

Let us look back to the equations for the barred variables in
terms of the unbarred ones (\ref{abgphibars})--(\ref{mbar}).  It
is apparent that, in order to generate a background with electric
charges from an uncharged one, we need to begin with a black hole
which has nonzero angular momentum (and thus a nonzero $G_{t
\varphi}$).  The higher--dimensional analogue of the
Schwarzschild solution is therefore insufficient for our
purposes.  Our starting point will instead be the
higher--dimensional analogue of the Kerr solution, which has mass
$M$ and just one\footnote{In $D$ dimensions there are
$[\frac{D-1}{2}]$ rotational Casimirs for a massive rep. of the
Poincar\'e group.}  rotation parameter $a$.  The metric is valid
for $D>3$ and, in the analogue of Boyer--Lindquist coordinates,
it is\cite{MyersPerry}
\begin{eqnarray}
d{\widehat S}^2&=&d{\widehat s}^2 \nonumber\\ &=& - dt^2 +
\sin^2\theta (r^2 + a^2) d\varphi^2 + \Delta (dt + a \sin^2
\theta d\varphi)^2 \nonumber\\ & & + \Psi dr^2 + \sigma^2
d\theta^2 + r^2 \cos^2 \theta d\Omega_{D-2}^2
\end{eqnarray}
where
\begin{equation}
\Delta = \frac{\widehat\mu}{r^{D-5}\sigma^2} \qquad \Psi =
\frac{r^{D-5}\sigma^2}{r^{D-5}(r^2+a^2)-\widehat\mu} \qquad
\sigma^2 = r^2 + a^2\cos^2\theta \quad .
\end{equation}
We have used hatted variables to distinguish this solution from
the one which we are about to obtain.  Notice also that the line
elements in the string and Einstein metrics, $d\widehat S^2$ and
$d\widehat s^2$ respectively, are equal because there is no
dilaton for the rotating black hole; there are no gauge fields
either.  In the above, the parameter $\widehat\mu$ is related to
the {\sc ADM} mass by
\begin{equation}
\widehat M_{ADM} = \frac{(D-2) A_{D-2}}{16 \pi G_N} \widehat\mu
\end{equation}
where $A_{D-2} = 2\pi^{\frac{D-1}{2}}/ \Gamma(\frac{D-1}{2})$ is
the area of a unit sphere in $(D-2)$ dimensions, and $G_N$ is the
Newton constant which we will hold fixed throughout.

In order to begin to form the charged background, we need the
barred variables (\ref{abgphibars})--(\ref{mbar}).  The nonzero
quantities are
\begin{eqnarray}
{\bar{\widehat A}}^{(k+1)}_\varphi &=&\frac{\Delta a}{(\Delta
-1)} \sin^2\theta \nonumber\\ {\bar{\widehat M}} &=&
\mbox{diag}(\I1,\frac{1}{\Delta-1},\Delta-1) \nonumber\\
{\bar{\widehat\Phi}} &=& -{\textstyle\frac{1}{2}}\log(1-\Delta)
\nonumber\\ {\bar{\widehat G}}_{ij} &=& \widehat G_{ij}
\nonumber\\ {\bar{\widehat G}}_{\varphi \varphi} &=& \sin^2\theta
\Biggl[ (r^2+a^2)-\frac{\Delta a^2 \sin^2 \theta}{(\Delta-1)}
\Biggr] \quad .
\end{eqnarray}

After transforming the barred--hatted variables via
(\ref{o1p17p}) to the barred variables denoting the charged
background, and untangling some of the relations
(\ref{abgphibars})--(\ref{mbar}) between barred and unbarred
variables, we obtain
\begin{eqnarray}
G_{tt}^{-1} &=& - 1 +(\cosh\alpha\cosh\beta)f^- +
{\textstyle\frac{1}{2}} (\cosh^2\alpha+\cosh^2\beta)(1+f^+)
\nonumber\\ f^\pm &=&
{\textstyle\frac{1}{2}}(\Delta-1\pm\frac{1}{\Delta-1}) \quad .
\end{eqnarray}

For the remainder of this work, we will be interested in
studying possible relations between nonrotating extremal black
holes and elementary string states.  We therefore set the
rotation parameter $a$ to zero now.  This could not have been
done before doing the boosting operation, because we would not
have been able to generate the full complement of charged
backgrounds.

After untangling of the remaining relations, the fields of the
charged black hole are found to be
\begin{eqnarray}
e^{-2\Phi} &=& 1 +
\frac{\widehat\mu}{\rho}(\cosh\alpha\cosh\beta-1)
+\frac{\widehat\mu^2}{4\rho^2}(\cosh\alpha-\cosh\beta)^2
\nonumber\\ dS^2 &=& \frac{-(\rho^2- \rho\widehat\mu) dt^2} {[
\rho^2 + \rho\widehat\mu(\cosh\alpha\cosh\beta-1)+
\widehat\mu^2(\cosh\alpha-\cosh\beta)^2/4 ]} \nonumber\\ & & +
\frac{\rho dr^2}{(\rho-\widehat\mu)} + r^2 d\Omega_{D-2}^2
\nonumber\\ A_{t,L}^{\alpha} &=& \frac{-n_L^{\alpha}}{\sqrt{2}}
\frac{\widehat\mu\sinh\alpha [
\widehat\mu(\cosh\alpha-\cosh\beta)/2+\rho\cosh\beta ]} {[ \rho^2
+ \rho\widehat\mu(\cosh\alpha\cosh\beta-1)+
\widehat\mu^2(\cosh\alpha-\cosh\beta)^2/4 ]} \nonumber\\
A_{t,R}^{\alpha} &=& \frac{-n_R^{\alpha}}{\sqrt{2}}
\frac{\widehat\mu\sinh\beta [
\widehat\mu(\cosh\beta-\cosh\alpha)/2+\rho\cosh\alpha ]} {[
\rho^2 + \rho\widehat\mu(\cosh\alpha\cosh\beta-1)+
\widehat\mu^2(\cosh\alpha-\cosh\beta)^2/4 ]} \nonumber\\ M&=&
\I1_k + \pmatrix{ P(r) {\vec{n}}_L{\vec{n}}_L^T & & Q(r)
{\vec{n}}_L{\vec{n}}_R^T \cr & & \cr Q(r)
{\vec{n}}_R{\vec{n}}_L^T & & P(r) {\vec{n}}_R{\vec{n}}_R^T \cr}
\end{eqnarray}
where
\begin{equation}
\rho \equiv r^{D-3}
\end{equation}
and
\begin{eqnarray}
P(r) &=& \frac{\widehat\mu^2\sinh^2\alpha \sinh^2\beta/2} {[
\rho^2 + \rho\widehat\mu(\cosh\alpha\cosh\beta-1)+
\widehat\mu^2(\cosh\alpha-\cosh\beta)^2/4 ]} \nonumber\\ Q(r) &=&
\frac{ -\sinh\alpha\sinh\beta\widehat\mu
[\rho+\widehat\mu(\cosh\alpha\cosh\beta-1)/2 ] } {[ \rho^2 +
\rho\widehat\mu(\cosh\alpha\cosh\beta-1)+
\widehat\mu^2(\cosh\alpha-\cosh\beta)^2/4 ]} \quad .
\end{eqnarray}
These functions $P,Q$ satisfy the relation $ P^2-Q^2 = -2P$.

The mass parameters of these black holes are, in the Einstein and
string metric respectively\footnote{For comparison, the parameter
$m$ of \cite{Sen3} is equal to our $\frac{1}{2}\widehat\mu$ in
$D=4$; note also that our vector field normalizations differ by a
factor of two and thus so do the gauge charges.},
\begin{eqnarray}
M_{ADM} &=& {\widehat M}_{ADM}
\frac{[1+(D-3)\cosh\alpha\cosh\beta]}{(D-2)} \nonumber\\
\mu_{string} &=& {\widehat\mu}\cosh\alpha\cosh\beta \quad .
\end{eqnarray}

The electric charges are defined as
\begin{equation}
Q_{(L,R)}^{\alpha} = \frac{1}{A_{D-2}} \int_{r\to\infty}
r^{D-2}d\Omega_{D-2} F_{rt(L,R)}^{\alpha}
\end{equation}
and eyeballing the vector potentials we get
\begin{eqnarray}\label{qleftqright}
Q_{L}^{\alpha} &=& \frac{n_L^{\alpha}}{\sqrt{2}}(D-3)\, \widehat\mu\,
\sinh\alpha\cosh\beta  \nonumber\\
Q_{R}^{\alpha} &=& \frac{n_R^{\alpha}}{\sqrt{2}}(D-3)\, \widehat\mu\,
\sinh\beta\cosh\alpha \quad .
\end{eqnarray}

The dilaton charge
\begin{equation}
\Xi = \frac{1}{A_{D-2}} \int_{r\to\infty} r^{D-2}d\Omega_{D-2}
\nabla_r ({\textstyle{\frac{1}{2}}}\Phi)
\end{equation}
is given by
\begin{equation}
\Xi = \frac{(D-3)}{4} \, \widehat\mu\, (\cosh\alpha\cosh\beta-1)
\quad .
\end{equation}

As they should, all of these parameters reduce to those of the
old black hole upon sending the boost parameters $\alpha,\beta$
to zero.


\section{Extremal black holes and supersymmetry}
\setcounter{equation}{0}

In this Section we will study a particular subclass of the black
hole backgrounds obtained above.  We will look at the extremal
cases which we term ``type $\cal{R}$''\footnote{To avoid
confusion with the Type {\sc II} string, we do not use the
terminology ``type {\sc II}'' of \cite{Sen3}.} and which are
obtained by taking the limit
\begin{equation}\label{mbetafixed}
\beta \to \infty \qquad {\widehat\mu} \to 0 \qquad \mu_0 \equiv
{\textstyle\frac{1}{2}} \widehat\mu\cosh\beta \quad \mbox{fixed}
\quad .
\end{equation}
There are two other extremal classes which could be considered.
The first, type $\cal{L}$, are obtained by switching $\alpha$ for
$\beta$ in the previous equation.  This gives extremal black
holes with a fixed relationship between mass and left--handed
charge, but as we are interested in supersymmetry which lives on
the right side of the string, we will not concern ourselves with
this class of extremal black holes further.  The other class of
extremal black holes is a subset of both type $\cal{R}$ and type
$\cal{L}$, which can be obtained as a subclass of the type
$\cal{R}$ black holes as follows:
\begin{equation}
\mu_0 \to 0 \qquad \cosh\alpha \to \infty \qquad \mu_1 \equiv
\mu_0\cosh\alpha \quad {\mbox{fixed}} \quad .
\end{equation}
The magnitudes of the left-- and right--handed charges are equal
in this case, and are fixed in relation to the mass.  All of
these extremal black holes have null singularities, i.e. the
horizon coincides with the singularity.

The fields of the type $\cal{R}$ black holes are
\begin{eqnarray}
dS^2 &=& -e^{2\Phi} dt^2 + dr^2 + r^2 d\Omega_{D-2}^2 \,
\nonumber\\ A_{t,L}^{\alpha} &=& n_L^{\alpha} \,
\frac{-\rho\sqrt{2}\mu_0\sinh\alpha} {\bigl[ \rho^2 + 2
\rho\mu_0\cosh\alpha+\mu_0^2\bigr]} \nonumber\\ A_{t,R}^{\alpha}
&=& n_R^{\alpha} \, \frac{-(\rho\cosh\alpha
+\mu_0){\sqrt{2}}\mu_0} {\bigl[ \rho^2 + 2
\rho\mu_0\cosh\alpha+\mu_0^2\bigr]} \nonumber\\ e^{-2\Phi}
&=& \frac{1}{\rho^2} {\bigl[ \rho^2 + 2
\rho\mu_0\cosh\alpha+\mu_0^2\bigr]} \nonumber\\ M &=& \I1_k +
\pmatrix{ P(r) {\vec{n}}_L{\vec{n}}_L^T & & Q(r)
{\vec{n}}_L{\vec{n}}_R^T \cr & & \cr Q(r)
{\vec{n}}_R{\vec{n}}_L^T & & P(r) {\vec{n}}_R{\vec{n}}_R^T \cr}
\end{eqnarray}
where
\begin{eqnarray}
P(r) &=& \frac{2\mu_0^2\sinh^2\alpha} {\bigl[ \rho^2 + 2
\rho\mu_0\cosh\alpha+\mu_0^2\bigr]} \nonumber\\ Q(r) &=&
\frac{-2\mu_0\sinh\alpha(\rho+\mu_0\cosh\alpha)} {\bigl[ \rho^2 +
2 \rho\mu_0\cosh\alpha+\mu_0^2\bigr]} \quad .
\end{eqnarray}

The charges on the type $\cal{R}$ black holes are seen to be
\begin{eqnarray}\label{typeIIparams}
M_{ADM} &=& \frac{A_{D-2}}{8\pi G_N}(D-3) \, \mu_0\cosh\alpha
 \nonumber\\ Q_L^{\alpha} &=& (D-3)\sqrt{2} \mu_0\sinh\alpha \,
 {n_L^{\alpha}} \nonumber\\ Q_R^{\alpha} &=& (D-3)\sqrt{2}
 \mu_0\cosh\alpha \, {n_R^{\alpha}} \nonumber\\ \Xi &=&
 \frac{(D-3)}{2} \mu_0 \cosh\alpha =
 \frac{1}{2\sqrt{2}}\big|{\vec Q}_R\big| \quad .
\end{eqnarray}
{}From this we find a relation between mass parameter and both
charges
\begin{equation}\label{mu0qrql}
\frac{\mu_{string}^2}{4\cosh^2\alpha} =
\mu_0^2 = \frac{1}{2(D-3)^2}
\biggl[{\vec{Q}}_R^2-{\vec{Q}}_L^2\biggr]
\end{equation}
and a relation between the {\sc ADM} mass and right--handed
charge
\begin{equation}\label{mADMqr}
M_{ADM}^2 = \Biggl[\frac{A_{D-2}}{16\pi G_N}\Biggr]^2
2\vec{Q}_R^2
\end{equation}
This last formula looks suspiciously like a supersymmetry
condition.

We would now like to find out whether the type $\cal{R}$ black
holes do indeed possess a supersymmetry.  Supersymmetry of
four--dimensional electric--magnetic black holes with
Kaluza--Klein charges has been established previously in
\cite{CveticYoum}.

It is easiest to start with the transformation rules written in
the sigma--model frame where they are the simplest.  One can
obtain the transformations in the Einstein variables by doing the
appropriate field redefinitions.  The ten dimensional
supersymmetry transformations are, for the gravinito, dilatino
and gaugino respectively,
\begin{eqnarray}\label{susy10}
\delta \tilde\psi_m &=& \tilde\nabla_m \tilde\epsilon -
{\textstyle{\frac{1}{8}}} \tilde H_{mnp} \tilde\Gamma^{np}
\tilde\epsilon \nonumber\\ \delta\tilde\chi &=&
{\textstyle{\frac{-1}{2}}} \tilde\Gamma^m\tilde\partial_m
\tilde\Phi\tilde\epsilon +{\textstyle{\frac{1}{12}}}\tilde
H_{mnp}\tilde\Gamma^{mnp}\tilde\epsilon \nonumber\\
\delta\tilde\lambda^{I} &=& {\textstyle{\frac{1}{2}}}
\tilde F^{I}_{mn}\tilde\Gamma^{mn}\tilde\epsilon
\end{eqnarray}
where $\tilde F^{I}$ are the gauge fields of the Yang--Mills
multiplet.  The supersymmetry variations of the bosonic fields,
which are proportional to the fermionic fields, are of course
zero because we are considering a bosonic background of the
theory.

Using the ten dimensional supersymmetry rules (\ref{susy10}) and
the dictionary of
\cite{MaharanaSchwarz,SenIJMP,Sen404,ScherkSchwarz} to relate ten
and $D$ dimensional fields we obtain the supersymmetry
transformations for the toroidally compactified theory in any
dimension $D$.  Firstly, we note that the $p$ internal gravitinos
and $16$ gauginos will combine to make $(16+p)$ modulinos, which
is precisely the correct number appropriate for the number of
bosonic moduli \cite{SUGRApapers}.  The next step is to untangle
the relations between the moduli matrix and the internal
components of the ten dimensional metric, antisymmetric tensor,
and gauge fields, using the basis change
(\ref{basisU}),(\ref{basisAM}) and the ten versus $D$ dimensional
field relations (\ref{tenvsD}).  The fields so obtained are then
used to get the supersymmetry variations of the fermions in the
toroidally compactified ``$D$ dimensional'' bosonic black hole
background in their full glory.  We will list here the fields
which are nonzero for type $\cal{R}$ black hole backgrounds.

Defining $\tau\equiv \sum_{a=1}^{p} n_R^a n_L^a$ and
$\delta\equiv 1- \sum_{I=1}^{16}n_L^I n_L^I$, we find that
\begin{eqnarray}\label{10fields}
e^{\tilde\Phi} &=& \frac{\rho}{R(\rho)} \nonumber\\ \tilde A^I_t
&=& \frac{\sqrt{2}\mu_0\sinh\alpha}{R(\rho)}\, n_L^I \nonumber\\
\tilde A^I_\alpha &=& \frac{-\sqrt{2}\mu_0\sinh\alpha}{R(\rho)}
\, n_L^I n_R^\alpha \nonumber\\ \tilde B_{\alpha\beta} &=&
\frac{-\mu_0\sinh\alpha}{R(\rho)} (n_R^\alpha n_L^\beta
-n_L^\alpha n_R^\beta) \nonumber\\ \tilde B_{t\beta} &=&
\frac{(\mu_0\cosh\alpha\, n_R^\beta + \mu_0\sinh\alpha\,
n_L^\beta)}{R(\rho)} \nonumber\\ \tilde G_{\alpha\beta} &=&
\delta_{\alpha\beta} +
(\delta-1)\frac{\mu_0^2\sinh^2\alpha}{R^2(r)}(n_R^\alpha
n_R^\beta) + \frac{-\mu_0\sinh\alpha}{R(\rho)} (n_R^\alpha
n_L^\beta +n_L^\alpha n_R^\beta) \nonumber\\ \tilde
G^{\alpha\beta} &=& \delta_{\alpha\beta} +
\frac{P(r)}{2}(n_R^\alpha n_R^\beta +n_L^\alpha n_L^\beta) +
\frac{-Q(r)}{2}(n_R^\alpha n_L^\beta +n_L^\alpha n_R^\beta)
\nonumber\\ \tilde E^\alpha_a &=& \delta^\alpha_a +
\biggl[-1+\frac{-Q(r)}{\sqrt{2P(r)}}\biggr]n_R^\alpha n_R^a +
\sqrt\frac{P(r)}{2} n_L^\alpha n_R^a \nonumber\\ \tilde
E^a_\alpha &=& \delta^a_\alpha + \frac{\mu_0\sinh\alpha}{R(\rho)}
\biggl[\biggl(-1+\frac{-Q(r)}{\sqrt{2P(r)}}\biggr)n_R^\alpha
n_R^a + \sqrt\frac{P(r)}{2} n_R^\alpha n_L^a \biggr] \nonumber\\
A^{[1]\alpha}_t &=& A_L n_L^\alpha - A_R n_R^\alpha \qquad
A^{[2]}_{\alpha t} = A_L \, n_L^\alpha + A_R \, n_R^\alpha \qquad
A^{[3]I}_{t} = \sqrt{2} A_L \, n_L^I \nonumber\\
& &
\end{eqnarray}
where
\begin{equation}
A_L = \frac{\rho\mu_0\sinh\alpha}{K(\rho)} \qquad A_R =
\frac{\mu_0(\rho\cosh\alpha+\mu_0)}{K(\rho)}
\end{equation}
where we have defined
\begin{eqnarray}
R(\rho) &=& \rho+\mu_0(\cosh\alpha+\tau\sinh\alpha) \nonumber\\
K(\rho) &=& \rho^2+2\rho\mu_0\cosh\alpha+\mu_0^2 \quad .
\end{eqnarray}
Note that $e^{-\tilde\Phi}$ is a harmonic function, although the
$D$ dimensional $e^{-\Phi}$ is {\it not} harmonic unless
$\alpha=0$.\footnote{This makes the form of multi--black hole
solutions hard to guess, and is a symptom of the fact that the
$D$ dimensional dilaton and gauge fields for a multi--black hole
are in fact built out of more than one harmonic function (see
also \cite{Kallosh,Behrndt}).}

The supersymmetry variations of the fermions become
\begin{eqnarray}
\delta\lambda^I &=& (\partial_i\rho)
\tilde\Gamma^{0i}\biggl[e^{-\Phi} \tilde F_{t\rho}^I(1+n_R^\alpha
A_t^{[1]\alpha}) + \tilde\Gamma^{0a} n_R^\alpha E^\alpha_a
{\tilde F}_{t\rho}^I \biggr]\tilde\epsilon \nonumber\\
\delta\tilde\psi_\alpha &=&
{\textstyle{\frac{1}{4}}}(\partial_i\rho) e^{-\Phi}
\tilde\Gamma^{0i} \biggl[\biggl(
-G_{\alpha\beta}F^{[1]\beta}_{t\rho}- \tilde H_{\alpha t\rho} +
A_t^{[1]\beta} \tilde H_{\alpha\beta\rho}\biggr) \nonumber\\
& & -
\tilde\Gamma^{0b}E^\beta_b e^{-\Phi} \biggl(\partial_\rho
G_{\alpha\beta}-\tilde H_{\alpha\beta\rho}\biggr) \biggr]
\tilde\epsilon \nonumber\\ \delta\tilde\psi_0 &=&
({\textstyle{\frac{1}{2}}}\partial_i\rho)
\tilde\Gamma^{0i}\biggl[ -\partial_\rho\Phi
+{\textstyle{\frac{1}{2}}} E^\alpha_a
e^{-\Phi}\tilde\Gamma^{0a}\biggl(
G_{\alpha\beta}F^{[1]\beta}_{t\rho} - \tilde H_{\alpha t\rho} +
A^{[1]\beta}_t \tilde
H_{\alpha\beta\rho}\biggr)\biggr]\tilde\epsilon \nonumber\\
\delta\chi &=& ({\textstyle{\frac{1}{2}}}\partial_i\rho)
\tilde\Gamma^i\biggl[-\partial_\rho\tilde\Phi + e^{-\Phi}
E^\alpha_a \tilde H_{t\alpha\rho}\tilde\Gamma^{0a} +
{\textstyle{\frac{1}{2}}}E^\alpha_a E^\beta_b \tilde
H_{\rho\alpha\beta} \tilde\Gamma^{ab}\biggr]\tilde\epsilon
\nonumber\\ \delta\tilde\psi_i &=& \partial_i\tilde\epsilon +
({\textstyle{\frac{1}{4}}}\partial_i\rho)E^\alpha_a
e^{-\Phi}\biggl[ -G_{\alpha\beta}F^{[1]\beta}_{t\rho}+ \tilde
H_{\alpha t\rho} -A^{[1]\beta}_t \tilde H_{\alpha\beta\rho}
\biggr] \tilde\Gamma^{0a}\tilde\epsilon \nonumber\\ &{}&
+({\textstyle{\frac{1}{8}}}\partial_i\rho)
\biggl[(E^\alpha_a\partial_\rho E_{\alpha b} - E^\alpha_b
\partial_\rho E_{\alpha a}) - E^\alpha_a E^\beta_b \tilde
H_{\alpha\beta\rho}\biggr] \tilde\Gamma^{ab}\tilde\epsilon
\end{eqnarray}

and using the fields (\ref{10fields}) we get
\begin{eqnarray}
\delta\tilde\lambda^I &=& (\partial_i\rho) n_L^I
\tilde\Gamma^{0i}
\frac{\sqrt{2}\mu_0\sinh\alpha}{R(\rho)\sqrt{K(\rho)}}
\biggl[1-n_R^a \tilde\Gamma_{0a}\biggr]\tilde\epsilon \nonumber\\
\delta\tilde\psi_\alpha &=&
({\textstyle{\frac{1}{2}}}\partial_i\rho) \tilde\Gamma^{0i}
\frac{\mu_0\sinh\alpha}{R(\rho)\sqrt{K(\rho)}} \biggl(n_L^\alpha
+ \frac{\mu_0\sinh\alpha(1-\delta)}{R(\rho)}\, n_R^\alpha \biggr)
\biggl[1-n_R^a\tilde\Gamma_{0a}\biggr] \tilde\epsilon \nonumber\\
\delta\psi_0 &=& ({\textstyle{\frac{1}{2}}}\partial_i\rho)
\tilde\Gamma^{0i}\, \frac{\mu_0(\rho\cosh\alpha+\mu_0)}{\rho
K(\rho)} \biggl[1-n_R^a\tilde\Gamma_{0a}\biggr]\tilde\epsilon
\nonumber\\ \delta\tilde\chi &=&
({\textstyle{\frac{1}{2}}}\partial_i\rho) \tilde\Gamma^i\biggl(
\frac{-\mu_0(\cosh\alpha+\tau\sinh\alpha)}{\rho R(\rho)} +
\frac{- n_L^b \tilde\Gamma^{0b}\mu_0\sinh\alpha}
{R(\rho)\sqrt{K(\rho)}}\biggr)
\biggl[1-n_R^a\tilde\Gamma_{0a}\biggr]\tilde\epsilon \nonumber\\
\delta\tilde\psi_i &=& \partial_i\tilde\epsilon -
{\textstyle{\frac{1}{2}}} \partial_i\Phi
\tilde\Gamma_{0a}n_R^a\tilde\epsilon \quad .
\end{eqnarray}

All but one of these equations may be satisfied by requiring that
$\tilde\epsilon$ satisfy the algebraic projection condition
\begin{equation}\label{projcond}
\biggl[ 1 - n_R^a\tilde\Gamma_{0a}\biggr] \tilde\epsilon = 0
\quad .
\end{equation}
Notice that this condition involves only the right--handed vector
$\vec n_R$, which makes sense as the supersymmetry lives on the
right side of the string.

Lastly, the projection condition (\ref{projcond}) yields
\begin{equation}
\delta\tilde\psi_i = \partial_i \tilde\epsilon -
{\textstyle{\frac{1}{2}}}\partial_i\Phi \tilde\epsilon
\end{equation}
which can be satisfied if
\begin{equation}
\tilde\epsilon = e^{\Phi/2}\tilde\epsilon^{(0)}
\end{equation}
where $\tilde\epsilon^{(0)}$ is a constant spinor satisfying
(\ref{projcond}).  This $\tilde\epsilon$ is the spinor
appropriate to the string metric, and it is independent of the
time and the internal coordinates, $(t,x^\alpha)$, and of $\vec
n_L$.  It depends on the vector $\vec n_R$ only through the
projection condition (\ref{projcond}).

The number of unbroken supersymmetry parameters is determined by
(\ref{projcond}).  We proceed by defining the quantities
\begin{equation}
\tilde P^\pm({\vec{n}}_R) = \textstyle\frac{1}{2}\biggl[ \I1 \pm
\tilde\Gamma_0 n_R^a\tilde\Gamma_a\biggr]
\end{equation}
which can easily be seen to be projectors by using the Dirac
algebra and the fact that $\vec n_R$ is a unit vector.  In
addition $\tilde P^\pm({\vec{n}}_R)$ commute with
$\tilde\Gamma_{11}$.  We form the new [chiral] spinor
combinations
\begin{equation}
\tilde\epsilon^\pm({\vec{n}}_R) = \tilde
P^\pm({\vec{n}}_R)\tilde\epsilon
\end{equation}
and note that supersymmetry then requires via the projection
condition (\ref{projcond}) that the $\tilde\epsilon^-$ vanish.
In the meantime, the $\tilde\epsilon^+$ are unconstrained, so we
see that the type ${\cal R}$ extremal black hole backgrounds
break {\it half} of the ten dimensional supersymmetries, leaving
eight unbroken supersymmetry parameters.

The number of unbroken supersymmetries in $D$ dimensions depends
on how the ten dimensional spinors break up into $D$ dimensional
spinors.  We take our cue for which extended supersymmetry is
appropriate in a given $D$ by matching the coset
$O(16+p,p)/[O(16+p)\otimes O(p)]$, which parametrizes the scalar
manifold, to those found for various supergravities in
\cite{SUGRApapers}.  We find that the appropriate theory is $N=4$
in $D=4,5$; $N=2$ in $D=6,7$; and $N=1$ in $D=8,9$.  In each
case, the group $O(p)$ is the automorphism group for the
supersymmetry algebra; respectively these are
$SU(4),USp(4),SU(2)\otimes
SU(2),USp(2),U(1),\I1$\cite{SUGRApapers}.  In $D=6$ the two
$SU(2)$'s reflect the chiral nature of the supersymmetry algebra.
We mention here that, despite appearances, it is possible to have
a central charge in the asymptotic supersymmetry algebra even for
the cases $D=8,9$ where the supersymmetry is only
$N=1$\cite{DGHR}.  (This happens because the spinors can be taken
to be [pseudo-]Majorana, and the $\Gamma^0$ matrix is symmetric
in this representation.)  For the other cases there are
$[\frac{N}{2}]$ central charges.

For the case $D=4$ an explicit representation of the
$\tilde\Gamma$ matrices is
\begin{equation}
(\tilde\Gamma^m,\tilde\Gamma^a) = (\Gamma^m\otimes
\I1,\Gamma_5\otimes \Pi^a)
\end{equation}
where $(\Gamma_5)^2=\I1$.  With a split of
$\tilde\epsilon=\epsilon\otimes\eta\equiv(\epsilon^A)$ , where
$A$ runs from $1\ldots N=4$, the projectors are $P^{\pm A}_{\quad
B} =\textstyle\frac{1}{2}(\I1^A_B\pm\Gamma_0\Gamma_5
n_R^a\Pi^{A}_{a\, B})$.  We therefore get $N=2$ supersymmetry,
with the central charges equal and proportional to the magnitude
of the electric charge.  We note that this agrees with the
analysis of \cite{KLOPP} for the appropriate $D=4$ single gauge
field subclass of the black holes presented here.

To summarize, we have found that {\it all} of the type $\cal{R}$
black holes in dimension $D=4\ldots 9$ preserve half of the
supersymmetries.  We expect that this will protect the relation
between the mass and the right--handed charge in all dimensions
$D=4\ldots 9$.


\section{Entropy and the stretched horizon}
\setcounter{equation}{0}

Here we will investigate the entropy and the stretched horizon of
extremal type $\cal{R}$ black holes.  We notice firstly that the
event horizon of a generic black hole of Section 2 is located
where $\rho=r^{D-3} = \widehat\mu$, i.e.
\begin{equation}
r_H = \widehat\mu^\frac{1}{D-3} \quad .
\end{equation}
To calculate the Hawking temperature, we change variables to $u,
\Theta$, defined by
\begin{eqnarray}
u &=& \frac{2}{(D-3)} \sqrt{r^{D-3} - r_H^{D-3}} \nonumber\\
\Theta &=& \frac{i(D-3)t}{r_H(\cosh\alpha+\cosh\beta)} \quad .
\end{eqnarray}
We obtain for the near--horizon string metric\footnote{The
conversion factor at the horizon from the string metric to the
Einstein metric is simply a constant, $\Gamma_H^{\gamma/2}$,
where $\sqrt{\Gamma_H}=\frac{1}{2}(\cosh\alpha+\cosh\beta).$} in
$(u,\Theta)$ variables
\begin{equation}\label{hzmetric}
dS^2 \sim  r_H^{5-D} (u^2 d\Theta^2 + du^2) + r_H^2 d\Omega_{D-2}^2
\end{equation}
which is Rindler space in the $(u,\Theta)$ directions.  We see
from the definition of $\Theta$ above, and the fact that $\Theta$
must be $2\pi$ periodic, that the Hawking temperature is given by
\begin{equation}\label{temperature}
T = \frac{(D-3)}{2\pi r_H(\cosh\alpha+\cosh\beta)} \quad .
\end{equation}

In the type $\cal{R}$ extremal limit, we see that the temperature
(\ref{temperature}) goes to zero, except for the special case
$D=4$ where it remains finite due to a scaling ``accident'' (see
equation (\ref{mbetafixed})).  However, this finite temperature
in $D=4$ may be unphysical, in view of the supersymmetry
properties derived in the previous Section (see also
\cite{KLOPP}).  We will take the temperature of all of the type
$\cal{R}$ black holes to be zero.

Let us next inspect the fields of these black hole backgrounds
for their behavior near the horizon.  We keep the Newton constant
fixed in our discussions; upon restoring the factors of the
string coupling at asymptotic infinity we have for the string
metric
\begin{equation}
dS^2 = \frac{-\rho^2 g_\infty^{2\gamma}dt^2} {[\rho^2 +
2\rho\mu_0\cosh\alpha+\mu_0^2]}+ g_\infty^{2\gamma}\bigl[dr^2 +
r^2 d\Omega_{D-2}^2\bigr]
\end{equation}
so for extremal black holes
\begin{equation}\label{proper}
\bar r= g_\infty^\gamma r
\end{equation}
measures proper distance from the horizon, in string metric.

On the horizon, we find that
\begin{eqnarray}
R_{tt} &\sim& {\bar r}^{2(D-4)} \nonumber\\ R &\sim&
\frac{1}{{\bar r}^2} \nonumber\\ e^{-2\Phi} &\sim&
\frac{\mu_0^2}{\rho^2} \nonumber\\ F^{\alpha}_{rt,L} &\sim&
-Q_L^{\alpha}\frac{r^{D-4}}{\mu_0^2} \nonumber\\
F^{\alpha}_{rt,R} &\sim& +Q_R^{\alpha}\frac{r^{D-4}}{\mu_0^2}
\quad .
\end{eqnarray}
We see that at string proper distance
\begin{equation}\label{esh}
{\bar r} = C \quad {\mbox{i.e.}} \quad r =
\frac{C}{g_\infty^{\gamma}}
\end{equation}
where $C$ is a pure number of order one, higher order corrections
to the action (\ref{stringaction}) will become important.  This
is the location of the {\it stringy} stretched horizon, and it is
independent of any of the parameters of the black hole.

Now we come to the entropy.  In order to calculate it for a black
hole, we use the method of Gibbons and Hawking, working along
similar lines to \cite{KOP}.  By examination of the field
equations corresponding to the action in the Einstein metric,
(\ref{einsteinaction}), we find that on shell the scalar
curvature $R_g$ cancels against the scalars $\Phi,M$.  We also
find that the on-shell action, including the necessary extrinsic
curvature term and with the numerical coefficient reinstated, can
be written as a surface term:
\begin{equation}
S=\frac{1}{8\pi G_N} \int_{\partial M} \Bigl[ (K-K_0)+
{\textstyle{\frac{1}{4}}} n_\mu \Bigl( \sqrt{-g} e^{-\gamma\Phi}
A_\nu^{\alpha}(LML)_{\alpha\beta}F^{\beta\mu\nu} \Bigr) \Bigr]
\quad .
\end{equation}
We obtain for the entropy the expected relation
\begin{equation}\label{entropy}
\Sigma = \frac{1}{4 G_N} R_E^{D-2} A_{D-2}
\end{equation}
where $G_N=M_{Pl}^{2-D}$ is the Newton constant and $R_E$ is the
radius of the local $(D-2)$-sphere in the Einstein metric.

For the extremal black holes which we are studying, the
singularity is null and the area of the event horizon is
classically zero.  However, we do not expect this zero area to
survive higher order corrections.  In other words, we will use
the entropy calculated at the {\it stringy} stretched horizon,
which is located at of order one string unit of proper distance
away from the event horizon as in (\ref{esh}).  With use of the
relation (\ref{stringeinstein}) between the string and Einstein
metrics, the Einstein radial variable is found to be
\begin{equation}
R^{D-2}_E(D) = [r_{S.H.}]\mu_0 \quad .
\end{equation}
The entropy for the type $\cal{R}$ extremal black holes,
calculated at the stringy stretched horizon, is thus
\begin{equation}
\Sigma_{S.H.} = \frac{A_{D-2}}{4 G_N} \frac{C}{g_\infty^\gamma}\,
\mu_0 \quad .
\end{equation}

In units where $\langle e^{\Phi}\rangle_\infty = g_\infty^2$,
electric charges scale as
\begin{equation}
{\vec Q}(g_\infty=1) = \frac{1}{g_\infty^{\gamma}} {\vec
Q}(g_\infty \ne 1)
\end{equation}
but $\mu_0$ stays fixed as it is proportional to the {\sc ADM}
mass which stays fixed with $G_N$.  The dilaton charge does not
scale either.  Using the relation (\ref{mu0qrql}) between the
mass parameter and the left and right charges
\begin{equation}
\mu_0 = \frac{1}{g_\infty^\gamma}\frac{1}{\sqrt{2}(D-3)}
\sqrt{{\vec{Q}}_R^2-{\vec{Q}}_L^2} \nonumber\\
\end{equation}
we obtain finally
\begin{equation}\label{bhentropy}
\Sigma_{S.H.} = \frac{1}{g_\infty^{2\gamma}}
\biggl[\frac{C A_{D-2}}{4\sqrt{2} G_N (D-3)} \biggr]
\sqrt{ {\vec{Q}}_R^2-{\vec{Q}}_L^2} \quad .
\end{equation}

We might ask at this stage what would happen if we were to take
into account quantum corrections to the fields.  It has been
argued previously \cite{Sen1} that quantum corrections would lead
only to a renormalization of the coefficient in front of the
$O(16+p,p)$--invariant combination of the charges.  In general,
the tree level action will be modified by higher order
corrections, which will lead to corrections to the solution and
to the supersymmetry transformations.  The $O(16+p,p)$ symmetry
and the solution--generating symmetry are expected to be valid to
all orders in $\alpha^\prime$ (see e.g. \cite{HassanSen}), but
the transformation laws to a given order may look different to
the tree level laws when written in terms of the fields like
$G_{\mu\nu}$ which appear in the low--energy effective action.
Therefore, in general, both of the quantities which feed into the
entropy, viz.  the coefficient of $d\Omega_{D-2}^2$ and the form
of $e^{-2\Phi}$ near the horizon, will be modified.  We expect by
dimension counting and $O(16+p,p)$ invariance (which is available
in some scheme) that the area term for the entropy will be
renormalized by a factor changing only the number $C/G_N$.  We
will absorb this factor into $C$.  Note that, besides the area
term, there will also be subleading corrections to the area term
for the entropy to which our calculations are not sensitive.

If we choose to keep the string scale fixed, and not the Newton
constant as we have done here, then we find that the black hole
entropy is proportional to $1/g_\infty^2$ in {\it all} dimensions
$D$.  This result for the entropy was explained [for a
non--extremal black hole] in $D=4$ in terms of genus zero stringy
configurations in \cite{S1,SU}.  There it was also found that, in
the limit of a very large mass black hole (where the subleading
terms are not present), the area term for the entropy would
receive corrections from higher genera which would serve only to
renormalize the Newton constant.

The next thing to calculate is the logarithm of the degeneracy of
the elementary (electrically charged) string state appropriate to
the type $\cal{R}$ black hole.  This correponds to the entropy of
a string state with given mass and charges.  The degeneracy
corresponds, of course, to the different possible ways of
distributing oscillators.

{}From the mass relations on the left and right sides of the
heterotic string, and the fact that the moduli matrix tends
asymptotically to the identity matrix, we know that
\begin{equation}
m^2 = g_\infty^2 \biggl[
\frac{\vec Q_R^2}{g_\infty^{4\gamma}} +
2(N_R-{\textstyle{\frac{1}{2}}})\biggr] =
g_\infty^2 \biggl[
\frac{\vec Q_L^2}{g_\infty^{4\gamma}} + 2(N_L-1) \biggr]
\end{equation}
In this formula, the mass of the string state is measured
in Einstein frame (in Planck units) so as to facilitate
comparison to the black hole mass; this gives rise to the
overall factor of $g_\infty^2$.
The factors of $g_\infty^{2\gamma}$ under the charges $Q$
arise because the field equations tell us that
the conserved charges are ``$Q/g_\infty^{2\gamma}$''.
(Note that there are no $\alpha^\prime$'s around; these have
all been converted into powers of $g_\infty$ and of $G_N$
which is fixed, via the relation
$m_{Pl}^{-2} = G_N^\gamma = g^2\alpha^\prime$ .)

The left and right charges are then related by
\begin{equation}
2(N_L-1)-2(N_R-{\textstyle{\frac{1}{2}}}) =
\frac{1}{g_\infty^{4\gamma}} \biggl[{\vec{Q}}_R^2 - {\vec{Q}}_L^2
\biggr]
\end{equation}
where $N_L,N_R$ are the contributions to the mass from the
oscillators excited on the left and right sides of the string.
For the electrically charged states which would correspond to the
type $\cal{R}$ black holes, supersymmetry demands that the
right--handed sector be in the ground state, i.e. $N_R =
{\textstyle\frac{1}{2}}$.

The level density $d_{ES}$ of elementary string states as a
function of $N_L$ is a standard formula; it goes at large $N_L$
as
\begin{equation}
d_{ES} \sim ({\mbox{power prefactors}}) e^{\sqrt{N_L}/T_H}
\end{equation}
where $T_H=1/(4\pi)$ is the Hagedorn temperature.
So we
find that at large $N_L$ the leading term is
\begin{equation}
\log(d_{ES}) \sim \frac{1}{g_\infty^{2\gamma}}
\frac{1}{\sqrt{2}T_H} \sqrt{ {\vec{Q}}_R^2-{\vec{Q}}_L^2 }
\quad .
\end{equation}

Putting together the black hole and string entropies we obtain
the result
\begin{equation}\label{theresult}
\Sigma_{S.H.} = \biggl[\frac{C T_{H}^{(0)}A_{D-2}}{4
G_N(D-3)}\biggr] \log(d_{ES}) \quad .
\end{equation}
Note in particular that the constant of proportionality between
the black hole entropy calculated at the stretched horizon and
the logarithm of the degeneracy of the elementary string states
is independent of the string coupling.  This is a direct
consequence of the fact that we took the stretched horizon to be
one {\it string} unit of proper distance away from the event
horizon.  If we wish to set the two entropies equal to each
other, then the numerical factor can be arranged with $C\sim 1$.

To finish, we note that in making the identification of the
supersymmetric black hole and string state entropies, we had {\it
no} freedom to soak up any factors of the mass by using a
redshift factor.  This is a phenomenon peculiar to extremal black
holes, and has its origins in the balance of forces which permits
the supersymmetry.  The static force between two of these objects
is zero, due to cancellation of attractive gravitational and
scalar forces by repulsive electromagnetic forces.  Thus any
redshift, which can also be thought of as a gravitational
dressing, would be cancelled out by dressings from
electromagnetic and scalar effects.


\section{Conclusion}
\setcounter{equation}{0}

In this paper we have studied the $D$ dimensional generalizations
of the electrically charged black holes of \cite{Sen1}.  We have
found that the entropy of a type $\cal{R}$ extremal black hole
with a given mass and charges, when calculated at the stringy
stretched horizon, is the same as the entropy of the
corresponding string state, up to a numerical factor of order $1$
\footnote{and up to subleading corrections which we have not
addressed here}.  That this correspondence works in dimensions
other than four is a satisfying consistency check.  A remaining
puzzle is to understand the precise nature of the stringy physics
which leads to the existence of the stretched horizon for
extremal black holes.

We have also seen that the type $\cal{R}$ extremal black holes
are supersymmetric, as was expected from the fact that they are
extremal and nonrotating and also from the conjectured
correspondence to string states.  The structure of the unbroken
target space supersymmetries, involving the projection condition
(\ref{projcond}) is an explicit example of a general phenomenon
found by Kallosh recently\cite{Kallosh}.  The fact that half of
the supersymmetries are unbroken is a result of the fact that
there is only electric charge.  If there were also magnetic
charge in four dimensions, then we expect that only one--quarter
of the supersymmetries would remain
unbroken\cite{KLOPP,CveticYoum}.

We regard these findings as additional support for the idea that
extremal black holes of toroidally compactified heterotic string
theory may be thought of as string states of the same theory.
Another consistency check to perform is to compare the results of
tree level scattering of the elementary states in string theory
with a semiclassical moduli space calculation of the same process
for the (extremal) black holes\cite{CMP}.

Some of the previous comparisons of the entropy of black holes
and string states have concentrated on nonextremal black holes.
These would correspond to string states with mass greater than
charge.  Both of these objects are unstable; one because of
Hawking radiation and the other because they would correspond to
ten dimensional massive states which are apt to decay (see,
e.g. \cite{Turoketal}).  Comparison of these two process is
likely to be difficult, not least due to problems in extracting
the widths of excited states which are not on the leading Regge
trajectory in string theory\cite{Sundborg}.  However, it would be
interesting to know if these two processes are related somehow.

While this work was in progress we were informed by M. Cveti\v c
that supersymmetry of four dimensional extremal dyonic black
holes of toroidally compactified heterotic string theory has been
proven very recently in \cite{CveticYoumNew}.


\section*{Acknowledgements}
The author would like to thank Eric Bergshoeff and John Schwarz
for a useful communication, and Renata Kallosh, Ashoke Sen and
Edward Witten for helpful discussions and suggestions and for
their support.  We would also like to thank Finn Larsen and
Juan Maldacena for helpful discussions.

This work was supported in part by National Science Foundation
grant PHY90-21984.


\section*{Appendix}

Here we list our notation and conventions.

Indices from the first part of the alphabet are internal (the
toroidal directions), and those from the last part are external.
We use small greek letters for curved indices, and small latin
letters for tangent space indices.  For the time coordinate, we
use $t$ for a curved index and $0$ for a tangent space index.
Where a distinction for gauge indices is useful, capital latin
indices are reserved for ``true'' gauge indices, indicating
charges or fields from the ten dimensional Yang--Mills multiplet.
All repeated indices are summed over.

The spacetime signature is ``mostly plus'', $(-,+,\ldots,+)$.
Antisymmetric combinations of gamma matrices are
$\Gamma^{[m_1,\ldots,m_n]}=\Gamma^{[m_1}\ldots\Gamma^{m_n]}$, and
antisymmetrization is done with weight one, e.g.
$\Gamma^{mn}=\textstyle\frac{1}{2}
(\Gamma^m\Gamma^n-\Gamma^n\Gamma^m)$.

Covariant derivatives on spinors are given by $\nabla_m \epsilon
= \partial_m \epsilon + {\textstyle\frac{1}{4}}
\omega_{mnp}\Gamma^{np} \epsilon$.  Any derivative acts only on
the object sitting immediately to its right.

Bars on fields denote the special combinations
(\ref{abgphibars})--(\ref{mbar}) which transform simply under the
solution--generating transformation of Section 1.  Hats denote
the fields of the original rotating uncharged black hole which
was used to generate electrically charged backgrounds.  Ten
dimensional fields are denoted by tildes.


\end{document}